\documentclass[notitlepage,a4paper,aps,prl,twocolumn,superscriptaddress,nofootinbib,groupedaddress]{revtex4-2}
\usepackage{graphicx}
\usepackage[colorlinks=true, pdfstartview=FitV, linkcolor=blue, citecolor=red, urlcolor=magenta]{hyperref}

\usepackage{dcolumn}
\usepackage{graphics}
\usepackage{amssymb}
\usepackage{amsmath}

\usepackage{url}
\usepackage{color}
\usepackage[utf8]{inputenc}
\usepackage{lipsum}

\usepackage{ulem}
\usepackage{tikz}
\usetikzlibrary{arrows,decorations.markings}

\newcommand{\rmd}{\textrm{d}}

\begin{document}
\newcommand{\ufrj}{\affiliation{UFRJ — Universidade Federal do Rio de Janeiro, Instituto de Física, Caixa Postal 68528, Rio de Janeiro, Brasil}}

\title{Unraveling the effect of rotation on the confinement/deconfinement transition of the
quark-gluon plasma}
\author{Nelson R. F. Braga}\email{braga@if.ufrj.br}
\ufrj
\author{Alexsandre L. Ferreira Jr}\email{alexsandrej@if.ufrj.br}
\ufrj

\date{\today}

\begin{abstract}
There is an apparent contradiction in the current literature about the effect of rotation in the quark-gluon plasma (QGP). 
While results from lattice QCD predict an increase in the confinement/deconfinement critical temperature, approximated calculations and effective models, including holographic ones, lead to the opposite result. Noncentral heavy ion collisions form QGPs with relativistic rotational velocities. Thereby, a great interest was drawn into the effect of rotation in strongly interacting matter. In this work, we show that the apparent contradiction is associated with the  choices of the observer considered in each case. We consider a holographic description of a rotating plasma using a Myers-Perry black hole. 
For a static observer, the result is that the confinement/deconfinement temperature decreases with the angular velocity,
while for an observer corotating with the plasma the opposite behavior is found,  in agreement with lattice calculations.

\end{abstract}


\maketitle

\section{Introduction}

Heavy ion collisions, produced at the Relativistic Heavy Ion Collider (RHIC) and at the Large Hadron Collider (LHC) led to the discovery of the quark-gluon plasma (QGP). The intensive research activity devoted in the last two decades to the properties of this strongly interacting matter, that behaves like a perfect fluid, resulted in 
important advances in the understanding of strong interactions
\cite{Shuryak:2008eq,Casalderrey-Solana:2011dxg, Busza:2018rrf}.

However, there are still many aspects that require a better understanding. One example is the phase structure of QCD, that depends not only on temperature and density but is also affected by the presence of magnetic fields and by rotation of the medium. 

Noncentral collisions generate a plasma rotating at relativistic angular velocities of the order of $\Omega\sim 10^{22}\,s^{-1}$, as measured through the polarization of vector mesons \cite{S17}. This finding boosted the investigation of relativistic rotating systems, with particular interest in understanding the effect of rotation in the confinement/deconfinement critical temperature. Currently, there is an apparent inconsistency about this important point. On the one hand, holographic models like \cite{CZ21,BFJ22,Y23,WF24,ZH23,BJ24} predict that the critical temperature decreases with the angular velocity. Such a behavior is in agreement with hadron resonance gas models \cite{FFH21}; perturbative approximations \cite{CFS22,CFS24}; and strong-coupling expansions \cite{FS25,WCHR25}.
On the other hand, lattice QCD (LQCD) results go in the opposite direction and present an increase in the critical temperature with angular velocity \cite{BKKR21,BKKR22,CGM23}. Some recent works 
using effective and/or holographic models found results consistent with the lattice ones \cite{MT23,CSDH25,CCDH25} and even nonmonotonic behaviors for the confinement critical  temperature \cite{GHZ23,J24} have been reported.  Some reasons for this discrepancy were speculated, such as problems in the analytical continuation from imaginary to real angular velocities \cite{CGM23,BCA24,C24}; and the absence of nonperturbative gluonic effects \cite{J24}.

In this letter we will show that the seemingly opposite behaviors 
of the critical temperature with angular velocity, can have  a simple interpretation. We present an example where the different results correspond to ``measurements" of temperature made by different observers, with a relative accelerated motion. Static frames, like in some effective models, find a decreasing confinement temperature, while for a comoving one, as in LQCD \cite{BKKR21,BKKR22,CGM23,Yamamoto:2013zwa}, the behavior is the opposite. There is no contradiction between the two types of results. Both observers will see a given rotating plasma in the same phase (confined or deconfined)  despite the different temperatures.

The effect of rotation will be described in this work following a holographic, or gauge gravity, approach. The confinement/deconfinement phase transition corresponds to a  Hawking-Page (HP) transition \cite{HP83} in the gravity side \cite{W98}. The deconfined phase is represented by a geometry containing a black hole. In order to incorporate rotation, a Myers-Perry black hole (MP BH)\cite{Myers:1986un}--a full spinning asymptotic anti-de-Sitter (AdS) solution of Einstein's field equations--will be considered.

%


\section{Myers-Perry black hole}

The MP BH is a full spinning solution of Einstein's field equations. As such, it incorporates the effects of inertial forces, in contrast with previous holographic models, in which the gravitational part consists of a boosted cylindrical black brane \cite{CZ21,BFJ22,Y23,WF24,ZH23}.  

In its most general version, alongside with a time translation symmetry, the MP BH has two $U(1)$ symmetries, associated with two independent conserved angular momenta, characterized by two parameters $(a,b)$. Here a simplified version with equal angular momenta, corresponding to $a=b$, is considered (see the review \cite{ACKW24} and references therein).  

The line element of the equal angular momenta MP BH (being refereed simply  as MP BH from now on) reads:
\begin{equation}
    \rmd s^2=\frac{\rmd r^2}{G(r)}+\frac{r^2}{L^2}\rmd s^2|_{bdy}+\frac{2\mu}{r^2}\bigg[\rmd t+\frac{a}{2}(\rmd\psi+\mathrm{cos}(\theta)\rmd\phi)\bigg]^2,
    \label{eMP_ds2}
\end{equation}
with the blackening factor,
\begin{equation}
    G(r)=-\frac{2\mu}{r^2}\left(1-\frac{a^2}{L^2}\right)+2a^2\frac{\mu}{r^4}+\frac{r^2}{L^2}+1,
    \label{blackfac}
\end{equation}
and the effective mass parameter $\mu$,
\begin{equation}
    \mu=\frac{M}{(1-a^2/L^2)^3}=\frac{r_+^4(L^2+r_+^2)}{2L^2r_+^2-2a^2(L^2+r_+^2)},
    \label{masspar}
\end{equation}
where $r_+$ is the Horizon radius, i.e., the higher positive root of $G(r)$. Further, the line element on the conformal boundary is given by
\begin{equation}
    \rmd s^2|_{bdy}=-\rmd t^2+\frac{L^2}{4}\bigg[\rmd\theta^2+\rmd\psi^2+\rmd \phi^2+2\,\mathrm{cos}(\theta)\rmd\psi\rmd\phi\bigg],
    \label{bdy_ds2}
\end{equation}
where we can see that the boundary has a compact topology $R\times S^3$. The spatial hypersurfaces are given by three-spheres in nonstandard coordinates.

Now, the spacetime is spanned by the coordinates $(t,r,\theta,\psi,\phi)$, with range
\begin{align}
    -\infty<\,t&<\infty,\\
    0\leq \,r&<-\infty,\\
    0\leq \,\theta&\leq\pi,\\0\leq\,\psi&<4\pi,\\0\leq\,\phi&<2\pi.
\end{align}
This choice of coordinates, first presented in \cite{MS08,M09}, is useful as it turns the boundary static and the rotation occurs solely along the $\psi$ direction. This is manifest when we calculate the horizon and boundary angular velocities, $(\Omega_\psi|_H,\Omega_\phi|_H)$ and $(\Omega_\psi|_{bdy},\Omega_\phi|_{bdy})$, respectively, that give
\begin{align}
    \Omega_\psi|_H=&-2a\left(\frac{1}{L^2}+\frac{1}{r_+^2}\right),&\hfill \Omega_\phi|_H=&0\\
    \Omega_\psi|_{bdy}=&0,& \hfill \Omega_\phi|_{bdy}=&0.
\label{ang_vel}
\end{align}
The horizon is generated by the killing vector field $\xi_H=\partial_{t}+\Omega_{\psi}|_H\partial_{\psi}$, with which one can calculate the temperature
\begin{equation}
    T=\beta^{-1}_{BH}=\frac{r^{2}_+L^2(2r^{2}_++L^2)-2a^2(r^2_++L^2)^2}{2\pi r^2_+L^3\sqrt{r^2_+L^2-a^2(r^2_++L^2)}}.
    \label{bhtemp}
\end{equation}
Further, the entropy, angular momenta, and total energy are given by \cite{PS05,GPP05}
\begin{align}
S&=\frac{4\pi^3 r^4_+L}{\kappa^2\sqrt{r^2_+L^2-a^2(r^2_++L^2)}} ,\\
    J&=-\frac{4a\pi^2\mu}{\kappa^2},\\
    E&=\frac{\pi\mu}{4L^2}(3L^2-a^2),
\end{align}
which are seen to satisfy the first law
\begin{equation}
    \rmd E=T\rmd S+\Omega \rmd J.
    \label{1law}
\end{equation}
The thermodynamic relevant angular velocity is $\Omega=\Omega_\psi|_{H}-\Omega_\psi|_{bdy}=\Omega_\psi|_H$, as the boundary is static.

The plasma lives on the conformal boundary that is a compact three-sphere, as discussed before. The actual QGP produced in heavy ion collisions lives in a flat spacetime. The usual procedure to approximate the compact space into a (hyper)plane is to consider an approximation of small angle and large radius, turning the BH into a black brane. For the present case, such approximation is given by the coordinate change
\begin{align}
    \frac{L}{2}(\phi-\pi)&=\epsilon x,\\
    \frac{L}{2}\mathrm{tan}(\theta-\pi/2)&=\epsilon y,\\
    \frac{L}{2}(\psi-2\pi)&=\epsilon z.
\end{align}
Alongside, $t\to\epsilon t$, $r\to\epsilon^{-1}r$, and a large horizon radius $r_{+}\to\epsilon^{-1}r_+$. Then, substituting in the line element and taking $\epsilon\to0$, we find
\begin{multline}
    \rmd s^{2}=\frac{r^2}{L^2}\left[\frac{L^4\rmd r^2}{r^4-r^4_{+}}-\rmd t^2+\rmd x^2+\rmd y^2+\rmd z^2\right.\\+\left.\frac{r^4_+}{r^4(1-a^2/L^2)}\bigg(\rmd t+\frac{a}{L}\rmd z\bigg)^2\right].
    \label{bBB_ds2}
\end{multline}
While now the boundary is a flat spacetime, what we find is just a black brane boosted in the $z$ direction \cite{MM20}. If one compactifies the $z$ direction in a cylinder, the solution is the same found in previous holographic models \cite{CZ21,BFJ22,Y23,WF24,ZH23}. Being connected with a static solution by a boost, noninertial effects are washed out in the approximation and the solution is not useful for our purposes. Therefore, we will consider in our analysis of the deconfinement transition the full solution having a compact boundary. However, our findings can be applied to the flow in a plane, either locally or by a stereographic projection, to be discussed in the final remarks.


\section{Hawking-Page geometrical description of confinement}
The AdS/CFT correspondence relates a weakly coupled asymptotic AdS gravitational theory with a strongly coupled conformal field theory (CFT) on its boundary. In the version used here, a five-dimensional asymptotic AdS spacetime will correspond to an $\mathcal{N}=4$ super-Yang-Mills (SYM) field theory in four dimensions. At finite temperatures, the supersymmetry is broken and the SYM becomes approximately a pure gauge theory, which displays several phenomena of interest present in QCD, including confinement \cite{W98}.

In this context, Witten proposed  that the confinement-deconfinement transition in the field theory is dual to a HP phase transition between a black hole and thermal AdS spacetimes--being the latter and the former dual to a confined and a deconfined phase, respectively. So, the HP approach offers a simple geometrical instrument to describe the phase transition of the QGP \cite{W98}. 
In order to account for the geometrical phase transition, one must treat spacetime as a thermodynamical system by assuming the existence of a gravitational partition function $\mathcal{Z}$, and perform a semiclassical saddle point approximation
\begin{equation}
    \mathcal{Z}(\beta)=\int\mathcal{D}g\,\mathrm{e}^{-I_{E}[g]}\approx\mathrm{e}^{-I_{E}|_{on-shell}},
\end{equation}
where $I_{E}$ is the Euclidean action, found through a wick rotation, $t\to i\tau$, in the original action. The gravitational action here is given by the Einstein-Hilbert term plus a negative cosmological constant, $\Lambda=-12/L^2$, and the Gibbons-York (GY) boundary contribution, so that
\begin{equation}
    I_E=\frac{1}{2\kappa^2}\ \int^\beta_0\rmd\tau \int\rmd^{4}x\sqrt{g_E}\,(R-\Lambda)+I_{GY}.
\end{equation}
In the following, the GY boundary term will be ignored due to the regularization scheme, to be explained later. Moreover, $\beta=1/T$ is the period of the compactified Euclidean time direction. 

In this semiclassical approximation, the dominant contribution to $\mathcal{Z}$ comes from the classical solution, and the \textit{on-shell} Euclidean action becomes the free energy of the system. The thermodynamically favored geometry is the one which contributes the most to the partition function, i.e., has a smaller free energy \cite{HP83}. Depending on the temperature, it can be either the black hole or the thermal AdS space. 

Along this line, to understand rotational effects in the critical temperature, we shall investigate the transition between the equal angular momenta MP BH presented in Eq. \eqref{eMP_ds2} and a thermal AdS spacetime. In addition to the wick rotation, one must change the rotation parameter as $a\to i\hat{a}$, to find the Euclidean MP BH metric. Moreover, the thermal AdS spacetime is just the Euclidean MP BH with mass and rotation parameters set to zero:
\begin{equation}
    \rmd s^2_{EAdS}=\left(\frac{r^2}{L^2}+1\right)\rmd\tau^2+\frac{\rmd r^2}{\frac{r^2}{L^2}+1}+r^2\rmd S_3^2.
\end{equation}
Further, to calculate the \textit{on-shell} action, notice that   $R=-20/L^2$ and $\sqrt{g}=r^3\mathrm{sin}(\theta)/2^3$ for both spacetimes. So that, the free-energy densities have the same form
\begin{equation}
    \mathcal{E}=\frac{I_{E}|_{on-shell}}{V_{bdy}}=\frac{4}{L^3\kappa^2}\int_0^{\beta}\rmd \tau\int_{r_0}^{\tilde R}\rmd r\, r^3,
\end{equation}
being $V_{bdy}=4\pi^2(L/2)^3$ the boundary spatial volume; $r_0$ the minimum value of the radial coordinate, and $\tilde{R}$ a UV cutoff. The difference between the densities of the two geometries is in the set of parameters $(\beta,r_0)$, that are equal to $(\beta_{AdS},0)$ and $(\beta_{BH},r_+)$ for the AdS and MP BH, respectively.

The \textit{on-shell} action is a divergent quantity and, consequently, must be regularized. For that, we subtract a pure AdS contribution from the MP BH action, a regularization scheme known as background subtraction method. Given this regularization, the GY term does not contribute, as the black hole's part decays rapidly at infinity and the remaining terms cancel each other. 

Importantly, in the background subtraction method the geometries must match in the asymptotic region $r=\tilde{R}\to\infty$. For that, we must require that the compact directions have the same circumference, in other words, we set \cite{GPP05}
\begin{equation}
    \beta_{AdS}=\frac{\sqrt{\hat{G}(\tilde{R})}}{\sqrt{\hat{G}(\tilde{R})|_{(\hat{\mu},\hat{a})=0}}}\beta_{BH}\approx\left[1-\frac{\hat{\mu} L^2}{\tilde{R}^4}\left(1+\frac{\hat{a}^2}{L^2}\right)\right]\beta_{BH},
\end{equation}
where $\hat{G}(r)$ and $\hat{\mu}$ are the previously defined quantities (\ref{blackfac}) and (\ref{masspar}), respectively, with $a\to i\hat{a}$. Hence, the resulting regularized free-energy density reads
\begin{align}
    \Delta \mathcal{E}&=\lim_{\tilde{R}\to\infty}\big(\mathcal{E}_{BH}-\mathcal{E}_{AdS}\big)\nonumber\\&=\frac{\beta_{BH}}{L^3\kappa^2}\left[L^2\hat{\mu}\left(1+\frac{\hat{a}^2}{L^2}\right)-r^{4}_{+}\right].
\end{align}
When $\Delta \mathcal{E}$ is positive, the AdS state is thermodynamically favored and the system is confined. On the contrary, for $\Delta\mathcal{E}<0$, the MP BH is stable and the boundary plasma is in a deconfined phase.

Thus, $\Delta\mathcal{E}=0$ defines a critical temperature $T_c$ in which the transition occurs. For a given rotation parameter, the temperature is defined by the horizon radius $r_{+}$. This allows the identification of a critical horizon radius $r_{c+}$, associated with the critical temperature, given by
\begin{equation}
    \frac{r^{2}_{c+}}{L^2}=\frac{1-\hat{a}^2/L^2}{1+\hat{a}^2/L^2}=\frac{1+a^2/L^2}{1-a^2/L^2},
\end{equation}
reducing to the Schwarzschild-AdS critical radius, $r_{c+}=L$, for $a\to0$ \cite{W98}. Moreover, causality imposes $a/L<1$, as we will shortly see. Thence, the region $a/L>1$, where $r_{c+}/L$ becomes imaginary, is excluded.

At this point, the rotation parameter is analytically continued back to real values $i\hat{a}\to a$. In the case of lattice and perturbative approaches, this procedure could be a source of problems \cite{CFS22,CGM23,BCA24,C24}. The issue appears in one-loop calculations, where poles in the real line forbid the analytic continuation of imaginary angular velocities back to real values \cite{CFS22}. In contrast, in the holographic approach considered here, thermodynamic quantities are computed in a semiclassical approximation on the gravitational side. The on-shell action and the resulting regularized free energy remain pole-free and the limit $i\hat{a}\to a$ can be taken without concern.

Finally, substituting the value of the critical radius into Eq. \eqref{bhtemp} temperature one finds
\begin{equation}
    \frac{T_c(a)}{T_c(0)}=\frac{1}{3}\left(\frac{3-a^2/L^2}{1+a^2/L^2}\right),
    \label{T em funcao de r+}
\end{equation}
where we introduced $T_c(0)=3/(2\pi L)$, the critical temperature in the absence of rotation. 

\begin{figure}[!h]
\includegraphics[width=0.48\textwidth]{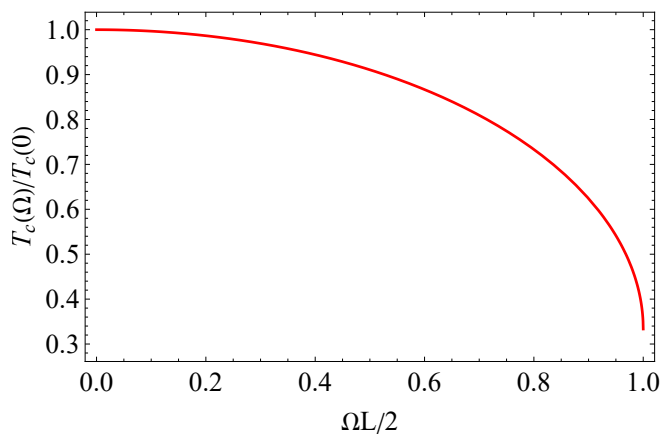}
\caption{Behavior of the critical temperature with respect to the angular velocity.}
\label{fig1}
\end{figure}

The angular velocity of the fluid is the one that enters the first law (\ref{1law}), and, therefore, can be obtained from  the free energy. Henceforth, it is given by the horizon angular velocity and depends not only on $a$, but also on the horizon radius. Therefore, at the critical radius $r_{c+}$, we have for the angular velocity
\begin{equation}
    \Omega=-2a\left(\frac{1}{r^2_{c+}}+\frac{1}{L^2}\right)=-\frac{4a/L^2}{1+a^2/L^2}.
    \label{angular velocity}
\end{equation}
Using Eqs. \eqref{T em funcao de r+} and \eqref{angular velocity} one finds the dependence of the critical temperature with $\Omega$, given by
\begin{equation}
    \frac{T_c(\Omega)}{T_c(0)}=\frac{2}{3\gamma}+\frac{1}{3},
    \label{tempcrit}
\end{equation}
where the Lorentz factor, $\gamma=1/\sqrt{1-(\Omega L/2)^2}$, appears naturally. Therefore, one can see that, as depicted in Fig. \ref{fig1}, the critical temperature decreases with rotation.

Being $L/2$ the radius of the three-sphere at the boundary, the linear velocity is given by
\begin{equation}
    v=\frac{\Omega L}{2}=-\frac{2a/L}{1+a^2/L^2}.
\end{equation}
 
Notice that, a luminal velocity, $v=1$, implies $a/L=1$, which translates into a divergent critical radius $r_{c+}$.

This behavior is similar to previous holographical models, in which the rotation decreases the confinement temperature \cite{BFJ22}. Note that the temperature in (\ref{tempcrit}) is measured by an observer that is not in the reference frame moving with the plasma.  In the following section, to compare our results with lattice QCD, we calculate the critical temperature as seen in a reference frame comoving with the plasma.


\section{Confinement/deconfinement temperature as measured by a corotating observer}

A naive expectation dictates that the temperature distribution in a thermodynamical system in equilibrium is homogeneous. However, it was demonstrated a long time ago that this statement does not hold in the presence of gravitational fields or, equivalently, of accelerations \cite{T30,TE30,SV19}. 

Take, as an example, a box filled with a gas of photons subjected to a downward constant gravitational acceleration. As photons climb the gravitational potential, they are red-shifted, losing energy and, consequently, being cooled. Thereby, in order for the box to have no diffusion of heat, the temperature must be greater at the bottom, where the gravitational potential is smaller \cite{SV19}. Consequently, the measured temperature will depend on the location of the observer.  

The same reasoning can be applied to systems under acceleration, such as a rotating plasma. The critical temperature calculated in the last section is measured in a reference frame that is not accelerated as the plasma. This can be understood noting that the boundary metric (\ref{bdy_ds2}) is static. In contrast, lattice calculations are performed in a reference frame corotating with the plasma \cite{BKKR21,BKKR22}. Therefore, to compare our results with LQCD we must calculate the critical temperature measured by a corotating thermometer, which we will call here the local temperature. 

\begin{figure}[!h]
\includegraphics[width=0.48\textwidth]{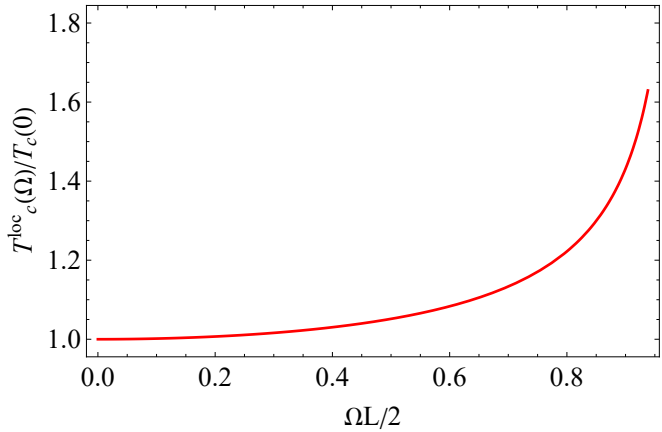}
\caption{Behavior of the local critical temperature with respect to the angular velocity.}
\label{fig2}
\end{figure}

With this purpose, we first transform to comoving coordinates, applying the change $\psi\to\psi+\Omega\, t$. The metric line element at the boundary (\ref{bdy_ds2}) becomes
\begin{align}
\rmd s_{loc}^2|_{bdy}=&-\frac{\rmd t^2}{\gamma^2}+\frac{L^2}{2}\Omega\,\rmd t\rmd \psi\nonumber\\&+\frac{L^2}{4}\bigg[\rmd\theta^2+\rmd\psi^2+\rmd \phi^2+2\,\mathrm{cos}(\theta)\rmd\psi\rmd\phi\bigg]\,.
    \label{cobdy_ds2}
\end{align}
The local temperature, $T^{loc}$, is then given through the Tolman-Ehrenfest law \cite{T30,TE30,SV19,Chernodub:2020qah}
\begin{equation}
    T^{loc}\sqrt{-g^{loc}_{tt}|_{bdy}}=\sqrt{-g_{tt}|_{bdy}}T\longrightarrow T^{loc}=\gamma T,
    \label{temperaturalocal}
\end{equation}
where $g^{loc}_{tt}|_{bdy}=-1/\gamma^2$ and $g_{tt}|_{bdy}=-1$ are the time components of the comoving and static boundary metrics, respectively. From Eqs. \eqref{tempcrit} and \eqref{temperaturalocal} one finds the  local critical temperature 
\begin{equation}
    \frac{T^{loc}_{c}(a)}{T_c(0)}=\frac{2+\gamma}{3}.
\end{equation}
This behavior is depicted in Fig. \ref{fig2}, and one can see that, for the local observer, confinement is disfavored by rotation because of noninertial effects.

This is a remarkable result indicating that the disagreement between LQCD and holographical models is just a consequence of the different observers considered in each case. While a corotating observer, as considered in LQCD,  finds that the critical temperature increases with angular velocity,  a nonrotating reference frame, as in the holographic models, observes a decrease of the critical temperature with the angular velocity. 

Nevertheless, it is important to stress that this does not mean that the same rotating plasma could be in the confined phase for one observer and in the deconfined for the other. The phase is the same for both observers. It is the temperature that is not the same, but rather related by \eqref{temperaturalocal}. The critical temperature is reached at the same rotational velocity for both observer, despite having different outcomes in each measurement process.
It is interesting to compare the results obtained here with ones found using LQCD. The latter are only valid for small values of the linear velocity $v=\Omega L/2$. Then, for comparison we perform an expansion in $v$ up to first order
\begin{equation}
    \frac{T^{loc}_c(v)}{T_c(0)}=1+B_2\,v^2,
\end{equation}
where for our case we have that $B_2=1/6\approx0.17$, while, lattice calculations lead to $B_2\approx0.7/1.3/0.5$, for open, periodic, and Dirichlet boundary conditions, respectively, with a weak dependence on lattice parameters \cite{BKKR21,BKKR22}.   
An exact quantitative consensus would not be expected, given that our model is based on the assumption of an approximate duality.  Despite that, our result is in the same order of LQCD when Dirichlet boundary conditions are imposed. 

Additionally, the three parameters, $\{r_+,a,L\}$, can be estimated by comparison with the QGP produced at the RHIC and the LHC. First, the measured critical temperature in the absence of rotation, $T_c(0)\sim155\,\,\mathrm{MeV}$ \cite{Niida:2021wut}, fixes the AdS radius to be $L=3.08\times10^{-3}\,\,(\mathrm{MeV})^{-1}$. For noncentral heavy ion collisions, indirect observations provide estimates of QGP temperature and angular velocities of $T\approx200\,\,\mathrm{MeV}$ and $\Omega\sim 7 \,\,\mathrm{MeV}$ \cite{S17}.  In our model, these parameters gives $v=\Omega L/2=0.012$ and critical temperature $T^{loc}_c(0.012)\approx T^{loc}_c(0)\approx155\,\,\textrm{MeV}$, which confirms that it is indeed in a plasma state.

Further, the given QGP measurements outcomes fix the remaining model parameters through Eqs. \eqref{ang_vel} and \eqref{bhtemp} to be: $r_+\approx5.02\times10^{-3}\,(\textrm{MeV})^{-1}$ and $a\approx2.42\times10^{-5}\,(\textrm{MeV})^{-1}$.

\section{Final remarks}

In this work it was shown that the disagreement between holographic models and lattice QCD, concerning rotational effects in the confined/deconfined transition, can be associated with the different choice of observers. Rotation of strongly interacting matter was holographically represented through a particular case of a MP BH \cite{Myers:1986un}. Then the measurement of temperatures of the different observers were obtained by considering two coordinate systems. One where the boundary is static and another where it is rotating. 
The local critical temperature, calculated in a corotating reference frame, increases with rotation, in agreement with lattice calculations. On the other hand, the inertial observer finds a critical temperature decreasing with rotation, as it was found using holographic models. This is a remarkable result indicating that effects of relative motion must be carefully taken into account when investigating rotation. Moreover, the qualitative agreement with LQCD, attests in favor of holography as a powerful auxiliary tool for studying strongly coupled systems such as the QGP. 

The relation between both measurement outcomes is given by the Tolman-Ehrenfest law, widely discussed in the context of the rotating QGP \cite{BKKR21,CGM23,C24}, either to certify that lattice results are found in the corotating frame or to discuss temperature distribution across the spatial extension of the plasma. Nonetheless, it was here applied for the first time, in the holography context, to discern between different observables. The well-established status of the Tolman-Ehrenfest law, a central piece in our argument, only attests in favor of the relevance of our findings.

At last, one might question the validity of our results for a QGP in a flat spacetime, as here the plasma is in a compact spatial hypersurface, a three-sphere. Nonetheless, it is well known that a sphere can be approximated locally to a flat space and, therefore, our local results will not change. To see that, perform a small angle approximation around $(\psi,\phi,\theta)\approx(0,0,\pi/2)$ at the boundary. With that, the boundary metric at leading order becomes

\begin{equation*}
    \rmd s^2|_{bdy}\approx-\rmd t^2+\frac{L^2}{4}\left(\rmd\theta^2+\rmd\psi^2+\rmd\phi^2\right),
\end{equation*}
which is just a flat space with $(x,y,z)=L/2(\phi,\psi,\theta)$.Nonetheless, such approximation is only valid for small distances as our estimates indicate a small value of the parameter $L$.

Additionally, a stereographic projection from $R\times S^3$ to $R^4$ was demonstrated to be a successful resource in describing the flux of a QGP \cite{G10}. More specifically, considering the MP BH, a projection in the conformal boundary results in a flow with various features expected in the QGP produced by noncentral collisions, denoted as Bantilan-Ishii-Romatschke flow \cite{BIR18}.

Hereby, our conclusions elucidate the source of an apparent discrepancy between holography and LQCD and, together with previous results \cite{ACKW24}, establishes the MP solution, along with the holographic principle, as a valuable resource in understanding the behavior of rotating strongly coupled systems.

\section*{Acknowledgments}

The authors are supported by FAPERJ -Fundaç\~ao Carlos Chagas Filho de
Amparo à Pesquisa do Estado do Rio de Janeiro. N.R.F.B also acknowledges the support of  CNPq - Conselho Nacional de Desenvolvimento Científico
e Tecnológico and CAPES - Coordenação de Aperfeiçoamento de Pessoal de Nível Superior.


\bibliography{ref}

\end{document}